\documentclass[aps,prl,twocolumn,amsmath,amssymb,groupedaddress,
longbibliography]{revtex4-1}

\usepackage{graphicx}
\usepackage{dcolumn}
\usepackage{bm}

\begin{document}
\title{Quadratic Optical Responses in a Chiral Magnet}
\author{Shun~Okumura$^1$, Takahiro~Morimoto$^2$, Yasuyuki~Kato$^2$, and Yukitoshi~Motome$^2$}
\affiliation{$^1$The Institute for Solid State Physics, The University of Tokyo, Kashiwa 277-8581, Japan\\
$^2$Department of Applied Physics, The University of Tokyo, Tokyo 113-8656, Japan}

\begin{abstract}
Chiral magnets, which break both spatial inversion and time reversal symmetries, carry a potential for quadratic optical responses.
Despite the possibility of enhanced and controlled responses through the magnetic degree of freedom, the systematic understanding remains yet to be developed. 
We here study nonlinear optical responses in a prototypical chiral magnetic state with a one-dimensional conical order by using the second-order response theory. 
We show that the photovoltaic effect and the second harmonic generation are induced by asymmetric modulation of the electronic band structure under the conical magnetic order, and the coefficients, including the sign, change drastically depending on the frequency of incident lights, the external magnetic field, and the strength of spin-charge coupling. 
We find that both effects can be enormously large compared to those in the conventional nonmagnetic materials.
Our results would pave the way for next-generation optical electronic devices, such as unconventional solar cells and optical sensors, based on chiral magnets. 
\end{abstract}

\maketitle
As represented by the development of the laser in 1960s, the research field of optical responses has been extended from linear to nonlinear effects~\cite{Boyd2003, Hanamura2007}.
In particular, quadratic optical responses have attracted much attentions in solid state physics.
The well-known example is a photo-induced electric current due to a photovoltaic effect (PVE) in the $p$-$n$ junction of semiconductors~\cite{Williams1960}, which has been applied to photonics devices such as a solar cell and an optical sensor.
The other example is a second harmonic generation (SHG); two photons with the same frequency generate a new photon with twice the frequency~\cite{Franken1961}, which has been used as a wavelength converter.
Such quadratic optical responses can occur in not only heterostructures but also bulk systems under breaking of spatial inversion symmetry, such as semiconductors~\cite{Sipe2000,Bergfeld2003,Moore2010}, ferroelectrics~\cite{Miller1964,Koch1975,Kraut1979,Baltz1981,Young2012_1,Young2012_2}, and topological materials~\cite{MorimotoSciAdv2016,Taguchi2016,Chan2017,Wu2017,Juan2017,Shreyas2018,Osterhoudt2019}.

\begin{figure}[t]
\centering
\includegraphics[width=\columnwidth,clip]{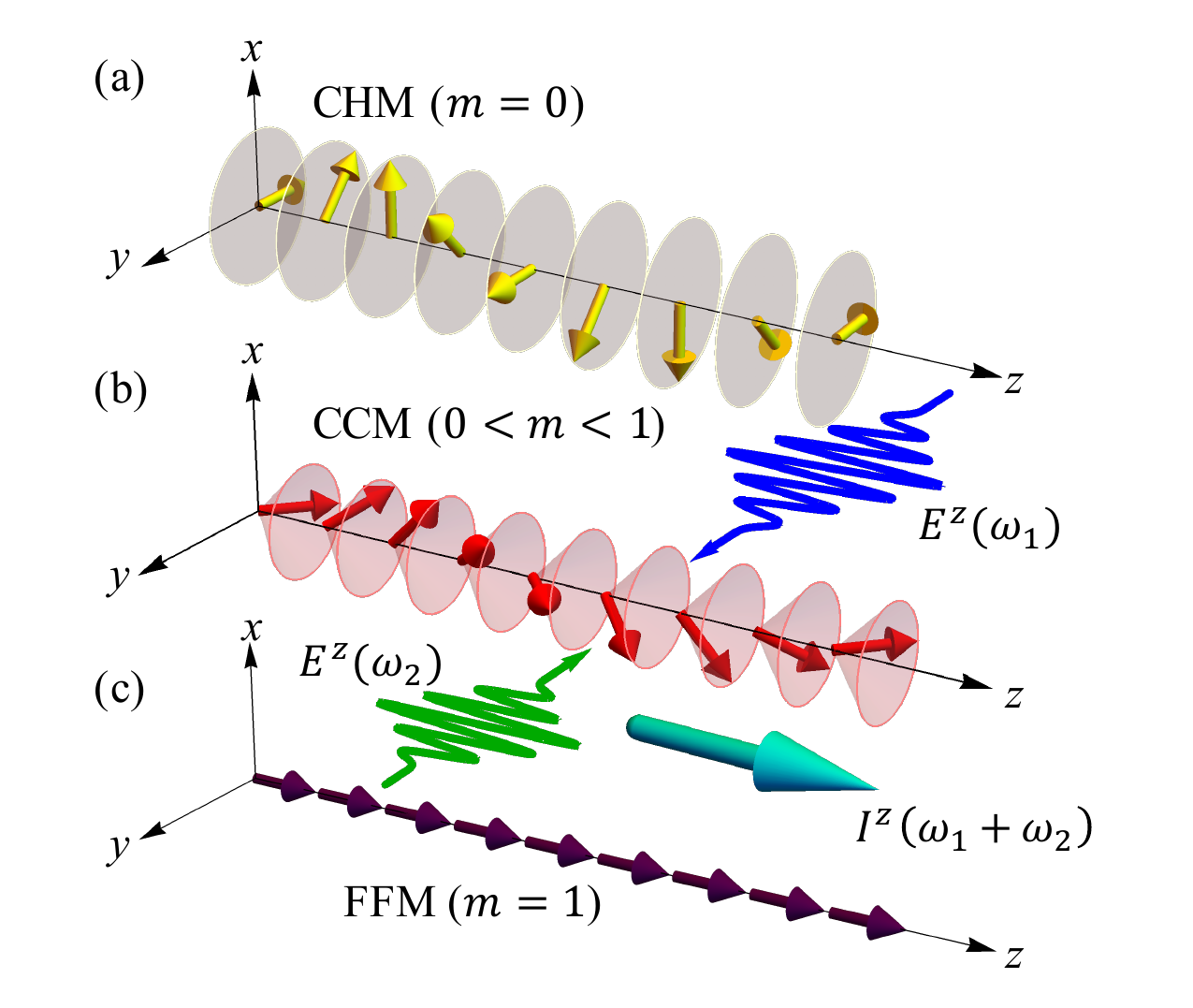}
\caption{
Schematic pictures of (a) a chiral helimagnetic state (CHM) at $m=0$, (b) a chiral conical magnetic state (CCM) at $0<m<1$, and (c) a forced ferromagnetic state (FFM) at $m=1$, where $m$ is the magnetization parallel to the $z$ axis.
The blue and green wavy arrows represent incoming linearly polarized lights oscillating in the $z$ direction with frequency $\omega_1$ and $\omega_2$, respectively.
The cyan arrow represents a nonlinear electric current generated in the CCM with frequency $\omega_1+\omega_2$ along the $z$ axis.
}
\label{f1}
\end{figure}

Recently, the search for the nonlinear optical responses has conducted for magnets since they potentially carry high controllability by an external magnetic field~\cite{Zhang2019,Watanabe2020,Watanabe2021}.
Amongst others, the magnets with noncentrosymmetric crystalline structures, called the chiral magnets, provide an excellent platform that meets the symmetry requirement. 
The chiral magnets are known to host peculiar magnetic textures, e.g., helices~\cite{Togawa2016}, skyrmions~\cite{Nagaosa2013,Kanazawa2017}, and hedgehogs~\cite{Fujishiro2020}, stabilized by the Dzyaloshinskii-Moriya (DM) interaction originating from the breaking of spatial inversion symmetry~\cite{Dzyaloshinsky1958,Moriya1960}.
Although the linear responses such as the topological Hall effect~\cite{Nagaosa2013} and the magnetoelectric effect~\cite{Tokura2010, Takahashi2012} have been understood based on the spin Berry mechanism~\cite{Nagaosa2012}, the systematic investigation of nonlinear ones remains yet to be developed despite the importance for future application to next-generation devices.

In this Letter, we theoretically investigate the quadratic optical responses in electrons coupled with a prototypical one-dimensional chiral magnet.
Our model exhibits magnetic textures changing from a chiral helimagnetic state (CHM) [Fig.~\ref{f1}(a)] to a chiral conical magnetic state (CCM) with spin canting [Fig.~\ref{f1}(b)], and to a forced ferromagnetic state (FFM) [Fig.~\ref{f1}(c)] while increasing the magnetic field applied to the helical axis (the $z$ axis in Fig.~\ref{f1}).
By using the second-order response theory, we show that the CCM, which breaks both spatial inversion and time reversal symmetries, induces the PVE and SHG through the asymmetric modulation of the electronic band structure.
We find that these quadratic responses vary drastically depending on the frequency of the incident lights, the magnetization, and the spin-charge coupling; 
in particular, the PVE can change its sign, meaning that the direction of the photo-induced current can be switched by these parameters.
Furthermore, we find that the PVE and SHG can be colossal in comparison with the conventional semiconductors and ferroelectric materials.
Our results would lay a cornerstone for the nonlinear transport and optical responses in various chiral magnets.

We consider a one-dimensional spin-charge coupled model, whose Hamiltonian is given by
\begin{eqnarray}
\mathcal{H} = &-&\sum\limits_{l,j,\mu}t_j(c^{\dagger}_{l\mu}c^{\;}_{l+j\mu}+\mathrm{h.c.})-J\sum\limits_{l,\mu,\nu}c^{\dagger}_{l\mu}{\boldsymbol \sigma}_{\mu\nu}c^{\;}_{l\nu}\cdot{\mathbf S}_{l}\nonumber\\
&-&\sum\limits_{l}\bold{D}\cdot(\bold{S}_{l}\times\bold{S}_{l+1})-\sum\limits_{l}\bold{h}\cdot\bold{S}_{l},
\label{eq:H}
\end{eqnarray}
where $c_{l\mu}(c^{\dagger}_{l\mu})$ is an annihilation (creation) operator for a $\mu$-spin electron at site $l$ on the periodic one-dimensional chain ($\mu = \uparrow$ or $\downarrow$), $\boldsymbol{\sigma} = (\sigma^x,\sigma^y,\sigma^z)$ are the Pauli matrices, and ${\mathbf S}_{l}$ is a three-component vector with normalized length $|{\mathbf S}_{l}|=1$ representing the localized classical spin at site $l$. 
The first term describes the kinetic energy of itinerant electrons, for which we take into account only the nearest-neighbor hopping $t_1$ and the next-nearest-neighbor hopping $t_2$ in the following calculations.
The second term represents the onsite coupling between the itinerant electrons and the localized classical spins with the coupling constant $J$.
The third term is the DM interaction with the DM vector $\bold{D}=D\hat{z}$, and the last term is the Zeeman coupling to the magnetic field along the $z$ axis, $\bold{h}=h\hat{z}$; $\hat{z}$ is the unit vector along the $z$ axis (see Fig.~\ref{f1}).

At zero magnetic field $h=0$, the ground state of this model is given by a CHM~\cite{Okumura2018PhysicaB}, and it is naturally expected that the spins are canted to the $z$ direction for $h>0$, resulting in the CCM. 
We therefore assume a chiral spin configuration given as 
\begin{eqnarray}
	\bold{S}_{l}=(\sqrt{1-m^2}\cos{Ql}, \sqrt{1-m^2}\sin{Ql}, m),
\label{eq:ansatz}
\end{eqnarray}
where $Q$ is the wave number specifying the period of the spiral spin structure and $m$ represents the magnetization per spin; both $Q$ and $m$ are determined by the model parameters in Eq.~\eqref{eq:H}.
The spin configuration describes the CHM at $m=0$ [Fig.~\ref{f1}(a)], the CCM for $0<m<1$ [Fig.~\ref{f1}(b)], and the FFM at $m=1$ [Fig.~\ref{f1}(c)].

By substituting Eq.~\eqref{eq:ansatz} to Eq.~\eqref{eq:H}, the Hamiltonian is reduced into a $2\times2$ matrix in the Fourier space up to a constant as
\begin{eqnarray}
	\mathcal{H} = &-&\sum\limits_{k',j,\mu}2t_j\cos\left(jk'-j\frac{Q}{2}\sigma^z_{\mu\mu}\right)c^{\dagger}_{k'\mu}c^{\;}_{k'\mu}\nonumber\\
&-&J\sum\limits_{k',\mu,\nu}\left(\sqrt{1-m^2}\sigma^x_{\mu\nu}+m\sigma^z_{\mu\nu}\right)c^{\dagger}_{k'\mu}c^{\;}_{k'\nu},
\label{eq:H_k}
\end{eqnarray}
where $k'=k+\frac{Q}{2}\sigma^z_{\mu\mu}$.
By the diagonalization, we obtain two energy bands split by $2J$, whose energy dispersions and eigenstates are denoted as $\varepsilon_{\pm}({k'})$ and $|\pm(k')\rangle$, respectively, where $+$($-$) represents the higher(lower)-energy band.

\begin{figure*}[t]
\centering
\includegraphics[width=\linewidth,clip]{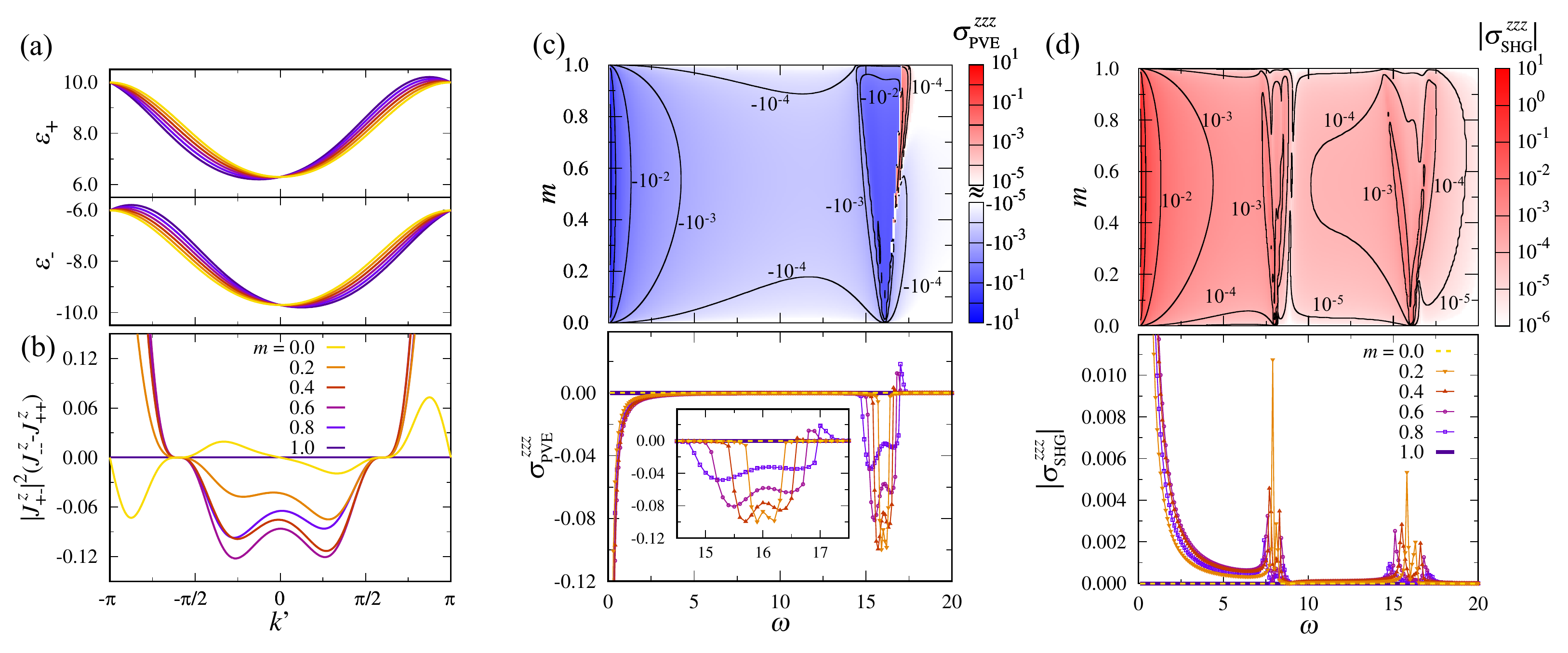}
\caption{
(a) Energy dispersions of itinerant electrons, $\varepsilon_{\pm}$, and (b) the coefficient in Eq.~\eqref{eq:injection}, $|J^{z}_{\pm}|^2(J^{z}_{--}-J^{z}_{++})$, as functions of $k'$ for several values of $m$.
The data are calculated for the strongly coupled case with $J=8.0$, $t_2=-0.1$, $D=0.12$, and $n=0.7$. 
Contour plots of (c) the photovoltaic coefficient $\sigma^{zzz}_\mathrm{PVE}$ in Eq.~\eqref{eq:sigma_PVE} and (d) the intensity of the SHG $|\sigma^{zzz}_\mathrm{SHG}|$ in Eq.~\eqref{eq:sigma_SHG} as functions of $\omega$ and $m$.
The lower panels show the $\omega$ dependences for several $m$.
}
\label{f2}
\end{figure*}

In order to investigate the nonlinear optical responses in this system, we calculate the optical conductivity by using the second-order response theory~\cite{Parker2019}.
For two incoming linearly polarized lights with frequencies $\omega_1$ and $\omega_2$ oscillating in the $z$ direction, denoted by $E^z(\omega_1)$ and $E^z(\omega_2)$, the nonlinear electric current is induced as 
\begin{eqnarray}
	I^z(\omega_1+\omega_2) = \sigma^{zzz}(\omega_1+\omega_2;\omega_1,\omega_2)E^z(\omega_1)E^z(\omega_2),
\label{eq:nonlinear_current}
\end{eqnarray} 
where the coefficient is the second-order optical conductivity given by 
\begin{eqnarray}
	&&\sigma^{zzz}(\omega_1+\omega_2; \omega_1, \omega_2)=\frac{1}{2N\omega_1\omega_2}\sum\limits_{k',a,b,c}\Bigg[f(\varepsilon_a)F^z_{aa}\nonumber\\
	&&+\frac{2f_{ab}J^z_{ab}T^z_{ba}}{\omega_1+i\gamma -\varepsilon_{ab}}+\frac{f_{ab}T^z_{ab}J^z_{ba}}{\omega_1+\omega_2+i\gamma -\varepsilon_{ab}}\nonumber\\
	&&+\frac{2J^z_{ab}J^z_{bc}J^z_{ca}}{\omega_1+\omega_2+i\gamma-\varepsilon_{ca}}\Bigg\{\frac{f_{ab}}{\omega_1+i\gamma-\varepsilon_{ba}}+\frac{f_{cb}}{\omega_1+i\gamma-\varepsilon_{cb}}\Bigg\}\Bigg]\nonumber\\
	&&+(\omega_1\leftrightarrow\omega_2).
\label{eq:second_order}
\end{eqnarray} 
See the schematic in Fig.~\ref{f1}.
Here, $N$ is the number of sites, $a$, $b$, $c=\pm$, $\varepsilon_{ab}=\varepsilon_a-\varepsilon_b$, $f_{ab}=f(\varepsilon_a)-f(\varepsilon_b)$, $J^z_{ab}=\langle a|\partial_k\mathcal{H}|b\rangle$, $T^z_{ab}=\langle a|\partial^2_{k}\mathcal{H}|b\rangle$, and $F^z_{ab}=\langle a|\partial^3_{k}\mathcal{H}|b\rangle$, $f(\varepsilon_{a})$ is the Fermi distribution function, and $\gamma$ represents the relaxation through the electron scattering, which is assumed to be a positive constant for simplicity.
In Eq.~\eqref{eq:second_order}, we take the elementary charge $e=1$, the reduced Planck constant $\hbar=1$, and the lattice constant $a_0=1$.

In the following, we focus on two interesting responses. 
One is the PVE which occurs for $\omega_1=-\omega_2=\omega$,
\begin{eqnarray}
\sigma^{zzz}_\mathrm{PVE}=\sigma^{zzz}(0; \omega,-\omega),
\label{eq:sigma_PVE}
\end{eqnarray} 
and the other is the SHG for $\omega_1=\omega_2=\omega$,
\begin{eqnarray}
\sigma^{zzz}_\mathrm{SHG}=\sigma^{zzz}(2\omega; \omega,\omega).
\label{eq:sigma_SHG}
\end{eqnarray} 
In general, $\sigma^{zzz}_\mathrm{PVE}$ is always real since the photo-induced current is a direct current, while $\sigma^{zzz}_\mathrm{SHG}$ is complex because the output is an alternating current. 
It is worth noting that, for the PVE, the fourth term in Eq.~\eqref{eq:second_order} with $a=c$ becomes dominant when $\gamma$ is small~\cite{SM1}; hence, the main contribution in Eq.~\eqref{eq:sigma_PVE} in the limit of $\gamma\to 0$ is written as
\begin{eqnarray}
	\frac{2\pi}{\gamma N\omega^2}\sum\limits_{k'}f_{+-}|J^z_{+-}|^2(J^z_{--}-J^z_{++})\delta(\omega-\varepsilon_{+-}),
\label{eq:injection}
\end{eqnarray} 
where $\delta(\omega)$ is the delta function.
This contribution is so-called injection current which originates from the group velocity of the excited carriers and is proportional to the relaxation time $\tau=\frac{1}{\gamma}$ in the steady state~\cite{Sipe2000}.
Hereafter, we use the coefficient $|J^z_{+-}|^2(J^z_{--}-J^z_{++})$ in Eq.~\eqref{eq:injection} as an indicator of the asymmetry of the energy bands.

In the following calculations, we set $t_1=1$ as the energy unit and we take $N=8192$ and $\gamma = 0.01$~\footnote{We need larger $N$ for smaller $\gamma$ to avoid an oscillation in $\sigma^{zzz}$ due to the finite-size effect. We checked the convergence of the results by varying $N$ for $\gamma=0.01$.}.
We study two cases:
One is the strongly coupled case where the spin-charge coupling $J$ is larger than the bandwidth of electrons and the two bands are split by the large $J$, and the other is the weakly coupled case where the splitting is small and the two bands overlap with each other. 
We take $J=8.0$, $t_2=-0.1$, $D=0.12$, and the electron filling $n=0.7$ for the former, while $J=0.2$, $t_2=-0.2$, $D=0.02$, and $n=0.3$ for the latter.
The parameter sets are chosen so that the CHM with $Q=\pi/4$ in Eq.~\eqref{eq:ansatz} is stabilized at zero field.
In the CCM for $h>0$, $m$ increases almost linearly with $h$ until $m$ is saturated in the FFM~\footnote{In the CCM, 
$Q$ slightly changes around the critical field where $m\to1$. In the present study, however, we fix $Q$ while varying $m$ in Eq.~\eqref{eq:ansatz} for simplicity.}.
We note that the sign of $\sigma^{zzz}$ is reversed when $Q$ ($m$) changes the sign, which is equivalent to the spatial inversion or mirror operation about the $xy$ plane (the $\pi$-rotation about the $x$ or $y$ axis).

\begin{figure*}[t]
\centering
\includegraphics[width=\linewidth,clip]{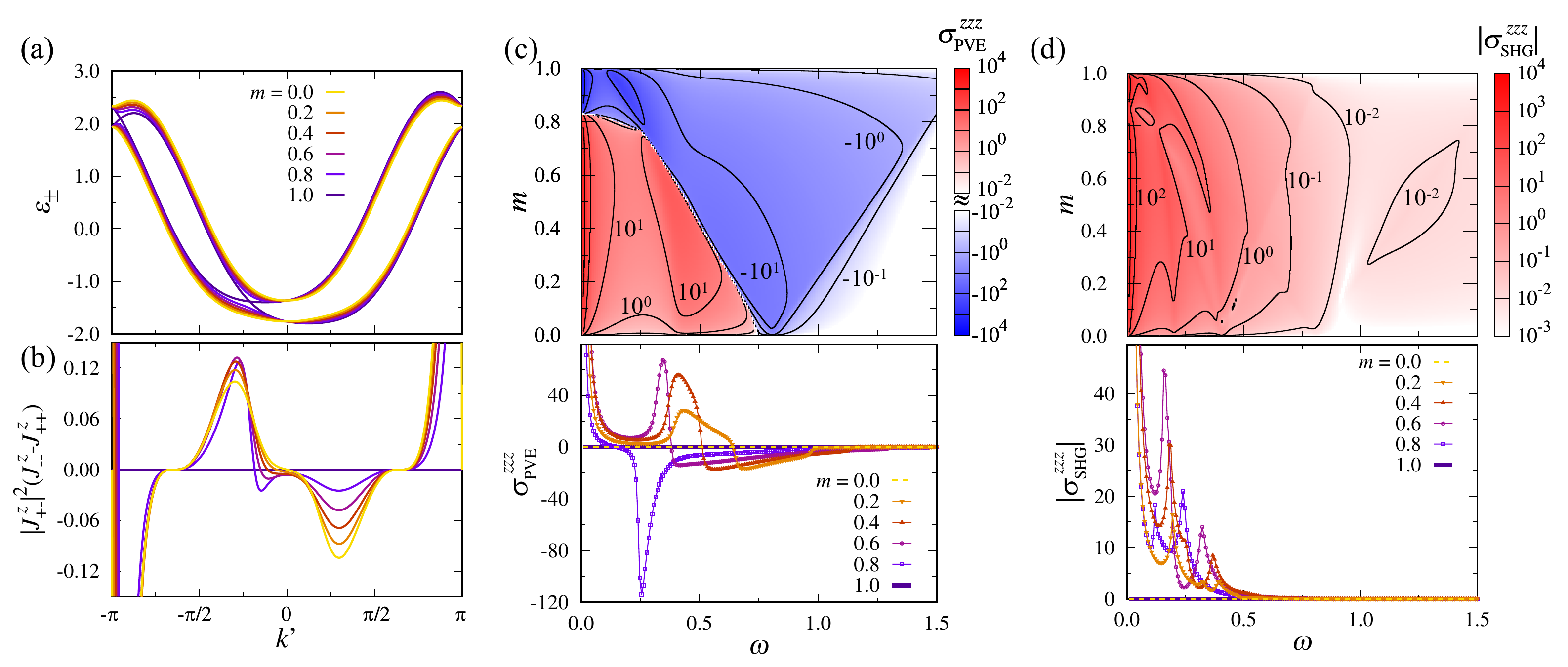}
\caption{
Similar plots to Fig.~\ref{f2} for the weakly coupled case with $J=0.2$, $t_2=-0.2$, $D=0.02$, and $n=0.3$. 
}
\label{f3}
\end{figure*}

First, we show the results for the strongly coupled case in Fig.~\ref{f2}.
In Fig.~\ref{f2}(a), we plot the energy dispersions $\varepsilon_{\pm}(k')$ for several $m$. 
The two bands are largely split by $\sim2J$.
While $\varepsilon_{\pm}(k')$ are symmetric with respect to $k'$ in the CHM at $m=0$, they show asymmetry in the CCM for $0<m<1$ where the spatial inversion and time reversal symmetries are both broken.
This asymmetry induces the quadratic optical responses.
In the FFM with $m=1$, $\varepsilon_{\pm}(k')$ recovers the symmetry with respect to the original wave number $k=k'\pm\frac{Q}{2}$. 
Figure~\ref{f2}(b) shows the coefficient in Eq.~\eqref{eq:injection}. 
While the coefficient is an odd function with respect to $k'$ in the CHM at $m=0$, it no longer is in the CCM for $0<m<1$. 
In the FFM at $m=1$, the coefficient vanishes. 
The results indicate that the injection current in Eq.~\eqref{eq:injection} is induced only in the CCM. 

We show the $m$ and $\omega$ dependences of the PVE coefficient $\sigma^{zzz}_\mathrm{PVE}$ in Fig.~\ref{f2}(c).
As expected from the above results, $\sigma^{zzz}_\mathrm{PVE}$ becomes nonzero in the CCM for $0<m<1$, while it vanishes in the CHM at $m=0$ and the FFM at $m=1$.
In the CCM, $\sigma^{zzz}_\mathrm{PVE}$ shows a sharp negative peak at $\omega\sim0$.
This is an intraband contribution from the coherent motion of the electrons, which leads to the nonreciprocal electric transport called the electric magnetochiral effect~\cite{Rikken2001,Yokouchi2017,Aoki2019,Jiang2020}.
This coherent peak diverges as $-\omega^{-2}$ due to the prefactor in Eq.~(\ref{eq:second_order}).
On the other hand, $\sigma^{zzz}_\mathrm{PVE}$ shows an ac response around $\omega\sim2J=16$, originating from the interband contributions. 
We find that this is dominated by the injection current in Eq.~\eqref{eq:injection}~\cite{SM1}.
Remarkably, $\sigma^{zzz}_\mathrm{PVE}$ exhibits the sign change in the large $m$ region.
This suggests that the direction of the photo-induced current can be switched by the external magnetic field. 

In Fig.~\ref{f2}(d), we also plot the $m$ and $\omega$ dependences of the intensity of the SHG, $|\sigma^{zzz}_\mathrm{SHG}|$.
Note that the argument of the complex $\sigma^{zzz}_\mathrm{SHG}$ gives just a phase delay of the output alternating current.
Similar to $\sigma^{zzz}_\mathrm{PVE}$, $|\sigma^{zzz}_\mathrm{SHG}|$ becomes nonzero only in the CCM for $0<m<1$. 
In contrast, however, it shows responses at $\omega\sim 0$, $J$($=8$), and $2J$($=16)$, 
which correspond to the intraband coherent motion, interband two-photon, and interband one-photon contributions, respectively.

Next, we show the results for the weakly coupled case in Fig.~\ref{f3}.
As shown in Figs.~\ref{f3}(a) and \ref{f3}(b), the energy bands and the coefficient of the injection current are asymmetrically distorted in the CCM for $0<m<1$, similar to the strongly coupled case.
While this also leads to the quadratic optical responses, we find that the behaviors can be more complicated because of the strong hybridization between the energetically overlapped bands.
As shown in Fig.~\ref{f3}(c), $\sigma^{zzz}_\mathrm{PVE}$ exhibits coherent and ac responses at $\omega\sim0$ and $2J$($=0.4)$, respectively, but the sign change occurs more drastically in wider ranges of $m$ and $\omega$ compared to the strongly coupled case in Fig.~\ref{f2}(c). 
In particular, the behavior at $\omega\sim0$ is relevant to the field-induced sign change of the electric magnetochiral effect in the CCM~\cite{Aoki2019}.
On the other hand, as shown in Fig.~\ref{f3}(d), $|\sigma^{zzz}_\mathrm{SHG}|$ shows multiple sharp peaks due to the overlap of the two- and one-photon contributions. 

Let us discuss the origin of the sign changes of $\sigma^{zzz}_\mathrm{PVE}$ in Figs.~\ref{f2}(c) and \ref{f3}(c).
In both strongly and weakly coupled cases, the ac responses are dominated by the contribution from the injection current in Eq.~(\ref{eq:injection}). 
This is given by the summation of the coefficient plotted in Figs.~\ref{f2}(b) and \ref{f3}(b) for $k'$ where the direct gap $\varepsilon_{+-}$ coincides with $\omega$; see Figs.~\ref{f2}(a) and \ref{f3}(a).
Thus, the sign changes in the optical regions are caused by the detailed balance under the asymmetrically modulated band structure.
On the other hand, the sign changes occur also in the coherent region at $\omega\sim0$ in the weakly coupled case, as shown in Fig.~\ref{f3}(c). 
In this case, they are caused by the competition among different contributions in Eq.~\eqref{eq:second_order}.
We note that such competition is not seen in the strongly coupled case, since all the contributions are negative for the parameters calculated here~\cite{SM1}.

Finally, we estimate the strength of the quadratic optical responses in our system. 
We assume the energy scale of the electron hopping $t_1\sim0.1$~eV and the relaxation time $\tau\sim1$~ps. 
Then, in the strongly coupled case, which mimics $d$-electron systems with the large Hund's coupling between the localized spins and itinerant electrons, the $\omega$ range of the ac response corresponds to $\sim10^2$~THz in the infra-red light region.
In this case, the magnitudes of PVE and SHG reach $\sim 10^{-4}$~$\mathrm{A/V^2}$ and $\sim 10^{-5}$~$\mathrm{A/V^2}$, respectively, which are comparable to those in the conventional ferroelectric material BaTiO$_3$~\cite{Young2012_1,Miller1964}.
On the other hand, in the weakly coupled case, which mimics the systems with the weak $s$-$d$ coupling, the ac response ranges from the transport region to the terahertz light or microwave region.
In this case, the PVE and SHG reach $\sim 10^{-2}$~$\mathrm{A/V^2}$ and $\sim 10^{-3}$~$\mathrm{A/V^2}$, respectively, which are two orders of magnitude larger than those in BaTiO$_3$ and comparable to those for topological materials such as TaAs~\cite{Osterhoudt2019,Wu2017}; these colossal responses originate from the small direct gap $\varepsilon_{+-}$.
Thus, our results indicate that the chiral magnets can generate unusually large quadratic optical responses, which can be controlled by the frequency of incident lights and external magnetic field as well as the electronic parameters; in particular, the large PVE coefficient can be controlled including its sign.

In summary, we investigated the quadratic optical responses, the PVE and SHG, in chiral magnets.
We found that the electronic band structure is modulated in an asymmetric way in the CCM which breaks both spatial inversion and time reversal symmetries, and it gives rise to the PVE and SHG whose magnitudes can be much larger than those in ferroelectric materials and comparable to topological materials.
In particular, we clarified that the PVE changes not only the magnitude but also the sign depending on the external magnetic field and the frequency of incoming lights.
This is a unique property of the chiral magnets where the band structure and the magnetism can be modulated by the magnetic field; in stark contrast, the PVEs in nonmagnetic systems, including the large PVE reported in Weyl semimetals~\cite{Osterhoudt2019}, do not show such controllability, and those in antiferromagnetic systems switch the sign only at the field-induced phase transition~\cite{Zhang2019}.
In realistic materials, the electronic structure would be more complicated, and furthermore, the electronic correlation may affect not only the velocity but also the lifetime of electrons~\cite{Cheng2015, Passos2018, Michishita2021}. 
Such complexity potentially gives rise to further nontrivial field and temperature dependence of the nonlinear optical responses.
Furthermore, while the present study is limited to the one-dimensional CCM, we expect that further nontrivial nonlinear responses can be observed in other chiral spin textures, for instance, magnetic skyrmion and hedgehog lattices.
\\

\begin{acknowledgments}
We would like to thank T.~Arima, H.~Ishizuka, N.~Kanazawa, Y.~Kato, J.~Kishine, Y.~Michishita, N.~Nagaosa, J.~Ohe, and Y.~Togawa for fruitful discussions.
This research was supported by JST CREST (No.~JPMJCR18T2 and No.~JPMJCR19T3), JST PRESTO (No.~JPMJPR19L9), the JSPS KAKENHI (No.~JP18K03447 and No.~JP19H05825), and the Chirality Research Center in Hiroshima University and JSPS Core-to-Core Program, Advanced Research Networks.
SO was supported by JSPS through the research fellowship for young scientists.
TM acknowledges funding from The University of Tokyo Excellent Young Researcher Program.
Parts of the numerical calculations were performed in the supercomputing systems in ISSP, the University of Tokyo.
\end{acknowledgments}

\appendix{
\section{Supplemental Material for "Quadratic Optical Responses in a Chiral Magnet"}
\subsection{Decomposition of the photovoltaic effect}

We discuss the PVE by evaluating each term in Eq.~\eqref{eq:second_order} separately as
\begin{eqnarray}
	&&\sigma^{zzz}_{\mathrm{PVE}, 1} = -\frac{1}{N\omega^2}\sum\limits_{k',a}f(\varepsilon_a)F^z_{aa},
\label{eq:decomposed_PVE1}
\\
	&&\sigma^{zzz}_{\mathrm{PVE}, 2} = -\frac{1}{N\omega^2}\sum\limits_{k',a,b}\left\{\frac{f_{ab}J^z_{ab}T^z_{ba}}{\omega+i\gamma -\varepsilon_{ab}}+\frac{f_{ab}J^z_{ab}T^z_{ba}}{-\omega+i\gamma -\varepsilon_{ab}}\right\},\nonumber\\
\label{eq:decomposed_PVE2}
\\
	&&\sigma^{zzz}_{\mathrm{PVE}, 3} = -\frac{1}{N\omega^2}\sum\limits_{k',a,b}\frac{f_{ab}T^z_{ab}J^z_{ba}}{i\gamma -\varepsilon_{ab}},
\label{eq:decomposed_PVE3}
\\
	&&\sigma^{zzz}_{\mathrm{PVE}, 4} = -\frac{1}{N\omega^2}\sum\limits_{k',a,b,c}\frac{2J^z_{ab}J^z_{bc}J^z_{ca}}{i\gamma-\varepsilon_{ca}}\Bigg\{\frac{f_{ab}}{\omega+i\gamma-\varepsilon_{ba}}\nonumber\\
	&&+\frac{f_{cb}}{\omega+i\gamma-\varepsilon_{cb}}+\frac{f_{ab}}{-\omega+i\gamma-\varepsilon_{ba}}+\frac{f_{cb}}{-\omega+i\gamma-\varepsilon_{cb}}\Bigg\}.
\label{eq:decomposed_PVE4}
\end{eqnarray}
Figures~\ref{f4} and \ref{f5} show the decomposed contributions as functions of $\omega$ and $m$ for the strongly and weakly coupled cases, respectively.
In both cases, $\sigma^{zzz}_{\mathrm{PVE}, 4}$ is dominant in the high-$\omega$ region, which is called the injection current in the limit of $\gamma\to 0$, while the contribution from other terms becomes comparable to $\sigma^{zzz}_{\mathrm{PVE}, 4}$ at $\omega\sim0$.
In the strongly coupled case, $\sigma^{zzz}_{\mathrm{PVE}, 2}$ in Fig.~\ref{f4}(b) and $\sigma^{zzz}_{\mathrm{PVE}, 4}$ in Fig.~\ref{f4}(d) change the sign in the high-$\omega$ region, while $\sigma^{zzz}_{\mathrm{PVE}, 1}$ in Fig.~\ref{f4}(a) and $\sigma^{zzz}_{\mathrm{PVE}, 3}$ in Fig.~\ref{f4}(c) are always negative.
On the other hand, in the weakly coupled case, the positive contributions of $\sigma^{zzz}_{\mathrm{PVE}, 2}$ in Fig.~\ref{f5}(b) and $\sigma^{zzz}_{\mathrm{PVE}, 3}$ in Fig.~\ref{f5}(c) are not negligible in the low-$\omega$ region in addition to the sign change of $\sigma^{zzz}_{\mathrm{PVE}, 4}$ in Fig.~\ref{f5}(d).

\begin{figure*}[t]
\centering
\includegraphics[width=\linewidth,clip]{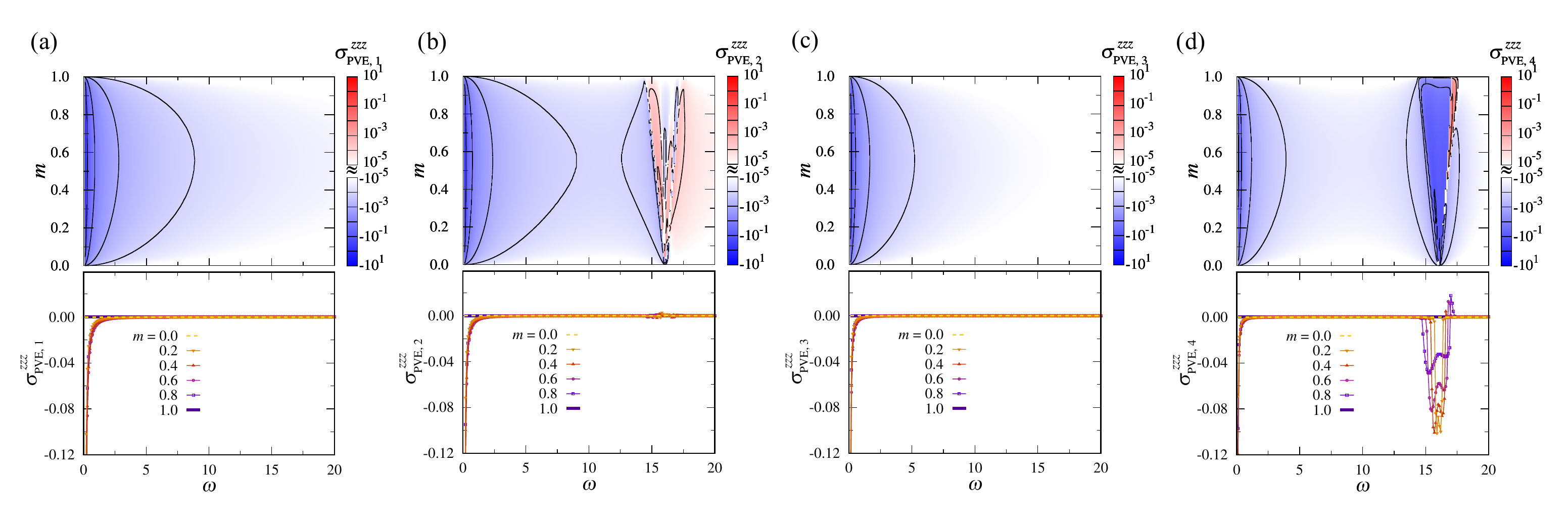}
\caption{
Contour plots of the decomposed photovoltaic coefficients $\sigma^{zzz}_{\mathrm{PVE,} i}$ in Eqs.~(\ref{eq:decomposed_PVE1})-(\ref{eq:decomposed_PVE4}) for $J=8.0$, $t_2=-0.1$, and $n=0.7$ as functions of $\omega$ and $m$.
The lower panels show $\omega$ dependences for several $m$.
}
\label{f4}
\end{figure*}

\begin{figure*}[t]
\centering
\includegraphics[width=\linewidth,clip]{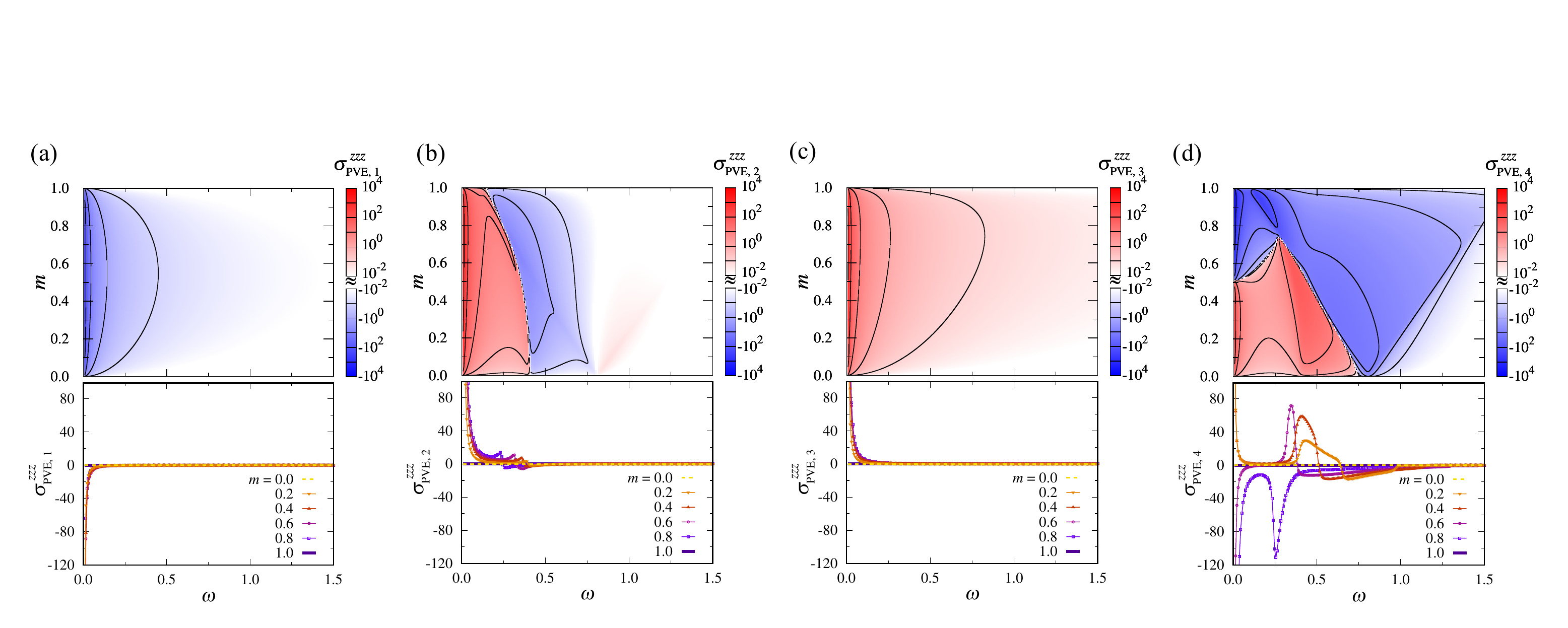}
\caption{
Similar plots to Fig.~\ref{f4} for $J=0.2$, $t_2=-0.2$, and $n=0.3$.
}
\label{f5}
\end{figure*}

}

\providecommand{\noopsort}[1]{}\providecommand{\singleletter}[1]{#1}%

\end{document}